# Multi–objective optimization of the dry electric discharge machining process

Sourabh K. Saha*, S.K. Choudhury

Department of Mechanical Engineering
Indian Institute of Technology Kanpur
Kanpur 208016, India

January 15, 2009

## Abstract

Dry Electric Discharge Machining (EDM) is an environment–friendly modification of the conventional EDM process, which is obtained by replacing the liquid dielectric by a gaseous medium. In this study, multi–objective optimization of dry EDM process has been done using the non dominated sorting genetic algorithm (NSGA II), with material removal rate (*MRR*) and surface roughness (*Ra*) as the objective functions. Experiments were conducted with air as dielectric to develop polynomial models of *MRR* and *Ra* in terms of the six input parameters: gap voltage, discharge current, pulse–on time, duty factor, air pressure and spindle speed. A Pareto–optimal front was then obtained using NSGA II. Analysis of the front was done to identify separate regions for finish and rough machining. Designed experiments were then conducted in these focused regions to verify the optimization results and to identify the region–specific characteristics of the process. Finishing conditions were obtained at low current, high pulse–on time and low duty factor, where as rough conditions were obtained at high current, low pulse–on time and high duty factor. Focused experiments revealed an additional process constraint based on the flushing efficiency.

*Corresponding Author: Sourabh Kumar Saha, email: sourabh.k.saha@gmail.com

# 1 Introduction

Electric Discharge Machining (EDM) is a thermo–electric process in which material removal takes place through the process of controlled spark generation. It is one of the most popular non-traditional machining processes being used today. EDM has achieved a status of being nearly indispensable in the industry because of its ability to machine any electrically conductive material irrespective of its mechanical strength. Despite its advantages, environmental concerns associated with the process have been a major drawback of EDM [1]. Hydrocarbon oil based dielectric fluids used in EDM are the primary source of pollution from the process. Replacing liquid dielectric by gases is an emerging field in the environment-friendly EDM technology [2–9].

Dry EDM is a modification of the conventional EDM process in which the liquid dielectric is replaced by a gaseous medium. Tubular tool electrodes are used and as the tool rotates, high velocity gas is supplied through it into the discharge gap. The flow of high velocity gas into the gap facilitates removal of debris and prevents excessive heating of the tool and workpiece at the discharge spots. Several experimental studies made in this field have brought out some of the essential features of the process. It is now known that apart from being an environment–friendly process, other advantages of the dry EDM process are: low tool wear, lower discharge gap, lower residual stresses, smaller white layer and smaller heat affected zone [3–6]. Several studies have also been made to improve the performance of the process [7– 9]. However, current literature is lacking in the area of dry EDM process optimization. In order to make dry EDM a cost effective and economically viable process, it is essential to have information on the optimum operating conditions.

A number of performance variables such as material removal rate ($MRR$), tool wear rate ($TWR$) and surface roughness ($Ra$) are commonly used to quantify the performance of an EDM operation. Seldom is anyone interested in only one of these parameters; more often than not it is desirable to have a high performance for more than one objective. In such cases selection of operating conditions becomes difficult because of conflicting objectives (such as $MRR$ and $Ra$), i. e. optimizing one objective leads to a poorer value for the other objective. One of the easier ways to tackle such multi-objective optimization problems is to convert them into single objective problems by attaching suitable weights to the individual objective functions. However, such techniques need predefined weight values for each objective function. Evolutionary multi-objective optimization (EMO) algorithms are population based stochastic methods, using which a well converged and uniformly distributed Pareto–optimal front can be obtained. Non–dominated Sorting GA (NSGA II) is one such EMO procedure which is commonly used for obtaining the Pareto–front. Several studies [10–12] have been made for multi-objective optimization in conventional EDM, however such studies are not available for dry EDM.

In the current work, multi-objective optimization of the dry EDM process has been performed. In order to optimize the process, first an empirical model of the process has been developed. Experiments were conducted based on the design of experiments (DOE) principles to obtain models of material removal rate (MRR) and surface roughness (Ra) in terms of the input variables. Multi-objective optimization was then performed using NSGA II, with MRR and Ra as the two objectives. The optimization results were then used to identify specific operating regions for performing finish and rough machining. Additional DOE based experiments were then performed in the focused regions to verify the optimization results and to improve the model.

## 2 Empirical Model Building

Experiments based on the Design of Experiments (DOE) principles were used to develop an empirical model of the dry EDM process. Experiments were conducted on a Z numerically controlled (NC) oil die-sinking EDM Machine, (R50 #ZNC, Elektra, Electronica Machine Tools India make). A dry EDM attachment unit was designed and developed to enable performing the dry EDM process on this machine [13]. The unit has been designed to fulfill the basic requirements of dry EDM: rotating tool and high velocity gas flow through the tubular tool. Process performance was quantified in terms of the volumetric material removal rate (MRR) and the center line average surface roughness value (Ra). The data obtained from a set of designed experiments based on Central Composite Design (CCD) runs was used to fit the models of MRR and Ra. Details of the experimental modeling process are discussed in a separate article [14] and summarized in this section.

### 2.1 Experimental Set-up

Dry EDM operations were performed on EN32 mild steel workpiece using tubular copper tool electrodes which had multiple non–central holes. No liquid dielectric was used during the experiments and the tool and work-piece were held in air medium. Pulsed DC power supply was provided to the electrodes, with the tool being at positive polarity and the workpiece at negative polarity. The dry EDM attachment unit was used to provide rotation to the tubular tool electrode while high velocity air was passed through it. A schematic of the dry EDM process is shown in Fig. 1. The flow of high velocity air through the tool electrode into the inter-electrode gap acts as the dielectric medium during dry EDM.

## 2.2 Procedure and Plan of Experiments

Blind holes were drilled on the workpiece during the experiments. Each dry EDM operation was performed for a fixed amount of time and the loss in weight of the workpiece was measured to obtain the volumetric material removal rate (*MR*R). The center line average (CLA) surface roughness parameter, *R*a was used to quantify the surface roughness. Surface roughness of the end surface of the holes was measured using a contact type stylus based surface roughness tester, Mitutoyo Surftest SJ-301. *R*a was measured along three different lines on the machined surface and the average value was considered for further analysis.

Based on the Central Composite Design (CCD), experiments were conducted to develop empirical models for *MR*R and *R*a, in terms of the six input variables: gap voltage ($V_g$), discharge current ($I_d$), pulse–on time ($T_{on}$), duty factor (*D*), air inlet pressure (P) and spindle rotational speed (*N*). Using CCD it is possible to fit models up to the second order including the quadratic terms. Each input variable (factor) was varied over five levels: ±1, 0 and ±α. The input variable values corresponding to each level are shown in Table 1. The total number of experimental runs was 86, with 64 factorial points, 12 axial points and 10 center points.

## 2.3 Regression Analysis

Design Expert 7.0.0 software was used for analyzing the experimental data. Values of various regression statistics were compared to identify the best fit model. The fitting was further improved by eliminating the insignificant terms through a backward step-wise model fitting. Regression statistics for the fitted models of *MR*R and Ra are shown in Table 2. The polynomial models for *MR*R and *R*a are shown below.

$$MRR = -1.04901 + 0.00536V_g + 0.02829I_d - 0.00044T_{on} + 0.00864D - 0.00226P - 0.0002N - 0.00066V_gI_d$$
$$+ 0.00003V_gT_{on} + 0.001I_dD + 0.00018I_dP + 0.00003I_dN - 0.00004T_{on}D \quad (1)$$

$$Ra = 7.95767 - 0.13375V_g - 0.01624I_d + 0.00334T_{on} + 0.00332D - 0.00211P - 0.00015N + 0.00038V_gI_d$$
$$- 0.000004T_{on}P + 0.00081V_g^2 - 0.000002T_{on}^2 \quad (2)$$

Where, MRR is in mm$^3$/min, $Ra$ in $\mu m$, $V_g$ in volt, $I_d$ in ampere, $T_{on}$ in $\mu s$, D in %, P in kPa and N in rpm.

## 3 Multi-objective Optimization

A multi-objective optimization problem arises when two or more objective functions are simultaneously optimized. Generally such problems consist of conflicting objectives so that it is not possible to obtain a single solution which is optimum in all the objectives. Instead of a single optimum solution, a set of optimum solutions called the Pareto–optimal set exists in such cases. Members of the Pareto–set are such that there exists no solution in the set which is better than the others in all the objectives; neither does a solution exist in the set which is worse than the others in all the objectives. By using evolutionary algorithms (such as GA), one can simultaneously obtain several Pareto–optimal points with an even distribution of the solutions. In the present work, multi–objective optimization of the dry EDM process has been done by considering maximization of MRR and minimization of Ra as two distinct objectives. The Non–dominated Sorting Genetic Algorithm (NSGA II) developed by Deb et al. [15] has been used to obtain the Pareto– optimal front of 1/MRR versus Ra.

### 3.1 NSGA II Algorithm

Non–dominated Sorting Genetic Algorithm (NSGA II) is a multi–objective evolutionary algorithm based on a fast non–dominated sorting principle [15]. The algorithm uses elitist non–dominated sorting along with crowding distance sorting to obtain the non–dominated set. The real parameter version uses the SBX crossover operator and polynomial mutation operator. The algorithm produces several non–dominated fronts after each generation. The non–dominated set which is obtained on convergence of the algorithm is very close to the Parteo–front.

The steps involved in the NSGA II algorithm are shown in Fig. 2 and briefly discussed here. First, a random population is initialized based on the variable ranges. The initial population is then sorted based on the non– domination criterion using a fast sorting method. This method not only utilizes the information about the the number of solutions that dominate the solution but also maintains the set of solutions which dominate it. This bookkeeping reduces the computational complexity of the sorting algorithm. The sorting classifies the population into several non–dominated fronts, each of which is assigned a rank.

Once the population is sorted, each member belonging to a front is assigned a crowding distance. The crowding distance of a solution is calculated as the Euclidean distance between the two neighboring solutions (on the same front) in the decision space. For this purpose, the solutions in each front are first sorted with respect to each objective function. The extreme members on a front are assigned a crowding distance value of infinity and the crowding distance of other members are normalized using the maximum and minimum objective function values. A high crowding distance implies that the corresponding member belongs to a less crowded region of the front and is important for maintaining diversity.

Selection of members is then carried out using the crowded–comparison operator. Although binary tournament selection is used, the selection criterion is modified to take into account the rank and crowding distance. While comparing, if the solutions belong to different fronts then the one with a lower rank is selected. However, if the solutions belong to the same front then the one with a higher crowding distance is selected.

The crossover and mutation genetic operators are applied on the selected population in a manner similar to that used during single objective GA. For real parameter implementations, simulated binary crossover (SBX) and polynomial mutation operator are used [16, 17]. $\eta$c and $\eta$m are the respective distribution indices for the SBX and polynomial mutation operators.

Finally, an elitist recombination strategy is used by combining the current population and the offspring population. For an initial population size of $N$, the combined population contains 2N members. The new population is obtained by picking members from each front successively until the size exceeds *N*. The members from the last added front are then sorted in descending order of crowding distance. A suitable number of members from this front are then picked so that a total of N members are obtained. All the steps starting from non–dominated sorting are repeated until the desired number of generations are completed.

## 3.2 Problem Formulation and Solution

Multi–objective optimization was done for simultaneously maximizing the *MR*R and minimizing the *R*a values. The available code was programmed for minimizing the objective functions [18]. Hence, 1*/MR*R and *R*a were considered as the two objective functions. The optimization problem can be represented as:

*Minimize 1/MRR; Minimize Ra*;

*Subjec*t *t*o : *MR*R > 0. (3)

Where, *MR*R and *R*a are given by Eq. 1 and Eq. 2 respectively. And, $V_g \varepsilon$ [55 99]V, $I_d$ ε [9 49]A, $T_{on}$ ε [50 1000]$\mu$s, D ε [8 88]%, P ε [58.8 245]kPa, N ε [300 2250]rpm.

The constraint *MR*R > 0 was added as the *MR*R was found to have negative values for certain combinations of the input parameters.

The NSGA II parameter values required for obtaining a converged and well distributed solution were obtained by hit and trial. NSGA II was run with a population size of 400. The input variables were represented in terms of real parameters. SBX distribution index of $\eta_c$ = 25 and polynomial mutation distribution index of $\eta_m$ = 150 was used. It was found that 100 generations were sufficient for convergence. Convergence was verified by comparing the objective function values at the Pareto–front extreme ends to the single objective optimized function values.

## 3.3 Results and Discussion

The non–dominated solutions obtained after the NSGA II run are shown in Fig. 3. The plot in Fig. 3 represents the Pareto–optimal front for minimization of the objective functions: 1*/MR*R and *R*a. *R*a is plotted on the x-axis and 1*/MR*R is on the y-axis. It can be seen that along the Pareto–front, for large values of *R*a low values of *MR*R were obtained (high 1/*MR*R value) and vice–versa; indicating that indeed *MR*R and *R*a are conflicting objectives for dry EDM optimization. All points along the front are "optimum in both the objectives" as defined by the Pareto–optimality criterion (since none of them is better than the other in both the objectives). This curve can be used to select an operating point when a target *R*a (or *MR*R) value is given. Several set of feasible operating conditions may exist for the same target *R*a (or *MR*R). But the best possible *MR*R for a given target *R*a (and vice–versa) can be directly obtained from the Pareto–optimal front. Thus, performance improvement can be obtained by selecting operating conditions along this curve. The extreme ends on the Pareto curve have special significance for machining. The end at which the tangent to the curve is perpendicular to the x-axis has the lowest Ra value, hence it represents the region suitable for finish machining. And the end at which the tangent to the curve is perpendicular to the y-axis has the highest MRR value and represents the region suitable for rough machining.

The input parameter values corresponding to the non–dominated solutions were also obtained by the NSGA II run. Out of a total of 400 solutions plotted in Fig. 3, a representative set of 15 solutions (and the corresponding input parameter values) is shown in Table 3. On inspection of the parameter values of the Pareto–optimal points it can be seen that all along the Pareto–front, air pressure and spindle speed have constant values (P = 245 kPa and N = 2250 rpm), which are the highest feasible value of the parameters. Thus, a high value of air pressure and spindle speed improves both the *MR*R and *R*a. This can be explained by the improved flushing efficiency at high values of P and N. A better removal of debris particles from the gap not only improves the *MR*R but also leads to a lower arcing probability. Arcing leads to surface damage, hence lower values of *R*a are obtained with low arcing probability. Forced flow of high velocity air into the discharge gap also helps in reducing the time required for recovery of dielectric strength of the medium since fresh and previously non–ionized medium is continuously supplied to the gap. This leads to higher process stability. Apart from improving the flushing efficiency, tool rotation also has an impact on the spark frequency. At high spindle speeds, a discharge may be interrupted even during the pulse–on time due to movement of the tool. Thus, several short sparks occur over a single pulse–on time at locations where the instantaneous inter–electrode gap is minimum; and the spark frequency increases with the spindle rpm. Since the same pulse energy is now distributed over a number of sparks, the crater depth is lower. Hence, lower Ra values are observed when the spindle speed is high. The values of other parameters ($V_g$, $I_d$, $T_{on}$ and $D$) vary along the Pareto–front. To analyze the trends in these variables we have to focus on specific regions of the front as has been done in Sec. 4 and 5.

When both the *MR*R and *R*a are important considerations for machining, a suitable operating point can be selected based on the information obtained from the Pareto–front. However, there may be instances when both the objectives may not be equally important. For example, during multi–pass operations machining may be performed in 3 steps: first a rough pass, followed by a semi–finish pass and finally a finish pass. For the semi-finish pass, an intermediate Pareto–optimal point (mid–way between the two extremes) may be a prudent choice. However for the rough and finish passes, regions near the respective ends are more suitable. Even in this region, operating near the Pareto front ensures a higher process performance. In the following sections (Sec. 4 and Sec. 5), separate finish and rough machining regions have been identified and focused experiments have been performed in these regions. The additional information obtained from these focused experiments not only allows to verify the optimization results, but also brings out the important features of the dry EDM process in these regions.

## 4 Finish Machining Region

### 4.1 Region Identification

From the results of the multi–objective optimization, separate regions for finish and rough machining conditions can be identified. The points which lie along the extreme end on the Pareto–front corresponding to the low *R*a values form the finish machining region. Out of the total points forming the Pareto–front, 2.5% of the points starting from the finish machining end were considered as forming the finish machining region and taken–up for further analysis. The Pareto–optimal points belonging to the finish machining region are shown in Table 4. The Pareto–optimal point having the minimum *R*a value has been highlighted in the table.

From Table 4 it can be seen that finish machining (low *R*a) can be obtained with low values of discharge current and duty factor. Low *R*a values are obtained due to a decrease in the spark energy under such conditions. Also, as noted in Sec.3.3 the highest values of air pressure and spindle speed were obtained for all the Pareto–optimal points. Additionally, optimization results show that an intermediate value of gap voltage exists for minimum $Ra$. Interestingly, high $T_{on}$ values were obtained for low $Ra$. It is generally considered that at high $T_{on}$ values spark frequency is low and hence high $Ra$ values should be obtained; since the spark energy is shared by a smaller number of sparks as opposed to the case when the spark frequency is high and a large number of sparks share the same energy. However, for the current dry EDM configuration, the spark frequency may be reasonably high even for high $T_{on}$ values due to tool rotation (as discussed in Sec. 3.3). The high $T_{on}$ values for finishing operation may be explained by the correspondingly high $T_{off}$ values (at low duty factors). A large pulse off–time favors efficient debris removal from the spark gap and low arcing probabilities, thus lowering the roughness value.

## 4.2 Focused Experiments

In the finish machining region, it was found that low duty factor and high values of pulse–on time, inlet air pressure and spindle speed were favorable for obtaining a low $Ra$ value (Table 4). The values of these parameters remain almost constant over the region, hence focused experiments were conducted keeping the values of these parameters constant. Since an optimum in voltage was found, a CCD design, capable of $2^{nd}$ order fitting, was chosen. Gap voltage and discharge current were varied in a small range around the optimum. Since values of current lower than 9 A (lower limit of current in original CCD experiment) were available on the EDM machine, the focused experiments were conducted at lower current values. Voltage was also considered as a factor in the CCD experiments because of the significant $V_g$/d interaction term in Eq. 2 [14]. The input parameter values and the output response ($MRR$ and $Ra$) for the CCD runs are shown in Table 5. Best $Ra$ value of 1.60 $\mu$m was obtained experimentally.

The experimental $Ra$ and $MRR$ values corresponding to the input conditions for the estimated minimum Ra (obtained by Pareto analysis) are highlighted in Table 5. It can be seen that the experimental $Ra$ value (2.21 $\mu$m) is higher than the estimated optimum (1.86 $\mu$m) by about 15.8%. Also, the corresponding experimental $MRR$ value (0.37 mm$^3$/min) is lower than the estimated optimum (0.89 mm$^3$/min) by about 140.5%. Thus, a poorer process performance was obtained experimentally for both $MRR$ and $Ra$. This indicates that the overall efficiency of the process is less than that estimated by the empirical model. One of the reasons for this could be that the effect of high inlet air pressure and high spindle rpm combination has been over–estimated by the model. High values of pressure and spindle speed lead to a higher flushing efficiency. However there exists a limiting value of flushing efficiency (100%), which is achieved when all the debris particles are flushed away from the gap. Since there is a limit to the flushing efficiency improvement, it may so happen that the flushing efficiency saturates at values lower than P = 245 kPa and N = 2250 rpm due to the combined effect of the two factors. Thus, the flushing efficiency estimated at the highest pressure and spindle speed combination is probably greater than 100% and hence not feasible in practice.

The effect of gap voltage and discharge current on $Ra$ is shown in Fig. 4(a) and Fig. 4(b) respectively. It was found that the $Ra$ value decreased with a decrease in the voltage and current indicating that lower pulse energy leads to a lower $Ra$ value. The interaction plot (Fig. 5) shows that there is no significant interaction effect between voltage and current in the finish machining region.

## 5 Rough Machining Region

### 5.1 Region Identification

In a manner similar to identification of the finish machining region (Sec. 4.1), the rough machining region was identified as the region near the Pareto–front end corresponding to high $MRR$ values (low $1/MRR$ values). 2.5% points starting from this end were considered to form the rough machining region. The corresponding Pareto–optimal points are shown in Table 6, with the point having the maximum $MRR$ highlighted in the table.

From Table 6 it can be seen that the maximum value of $MRR$ was obtained when the values of current, duty factor, inlet air pressure and spindle speed were highest. As discussed earlier in Sec. 4.1, spark energy increases with current and duty factor and the flushing efficiency is improved by high air inlet pressure and high spindle speed. Hence, $MRR$ is highest corresponding to these parameter values. Additionally, low values of $T_{on}$ were favorable for a high $MRR$ because the corresponding $T_{off}$ values were also low as the duty factor was high (88%). Due to a small non–cutting time, MRR is high under such $T_{on}$-$T_{off}$ conditions. It is interesting to note that for maximum MRR, low values of gap voltage were obtained. Spark energy increases with an increase in $V_g$, which is expected to increase the MRR. However, during dry EDM an additional effect of $V_g$ is observed. It is known that the discharge inter-electrode gap distance is directly proportional to $V_g$. Hence, at low $V_g$ the distance between tool end and workpiece is low. Thus, the effective air velocity striking the discharge gap is high as compared to the case when the tool is far away from the workpiece. This increases the flushing efficiency and the MRR is greatly improved due to an efficient debris removal from the discharge gap.

## 5.2 Focused Experiments

In the rough machining region it was found that low values of $V_g$ and $T_{on}$ and high values of $I_d$, $D$, P , and N were favorable for obtaining a high MRR. Since values of gap voltage lower than 55 V and pulse–on time lower than 50 µs were available on the EDM machine, experiments were conducted at lower $V_g$ and $T_{on}$ values. A $2^2$ factorial experiment was conducted with $V_g$ and $T_{on}$ as the factors. Three center runs were also included to check for curvature effect in the region. $I_d$, $D$, P , and N were held constant during the factorial runs. The input parameter values and the output response (MRR and Ra) for the factorial runs are shown in Table 7. Best MRR value of 6.83 mm$^3$/min was obtained experimentally.

The experimental MRR and Ra values corresponding to the input conditions for the estimated maximum MRR are highlighted in Table 7. It can be seen that the experimental MRR value (6.83 mm$^3$/min) is lower than the estimated optimum (8.18 mm$^3$/min) by about 19.8%. Also, the corresponding experimental Ra value (2.97 µm) is higher than the estimated optimum (2.84 µm) by about 4.4%. Thus, the model over–estimates the process efficiency even in the rough machining region similar to that in the finish machining region. However, the extent of overestimation is lower than in the finishing region. This could be due to a higher amount of debris formation during rough machining, because of which better flushing is required in the rough machining region. Thus, the saturation of flushing efficiency takes place at higher values of inlet air pressure and spindle speed. At the highest pressure and spindle speed combination there is indeed an increase in flushing efficiency as predicted by the model. The empirical model (Eq. 1 and Eq. 2) is thus closer to experimental values in the rough machining region than in the finish machining region.

The effect of gap voltage and pulse–on time on *MR*R is shown in Fig. 6(a) and Fig. 6(b) respectively. It can be seen that the *MR*R increased when either $V_g$ or $T_{on}$ was increased. However, the central runs indicate that there is sufficient curvature in the region. The *MR*R values of central points lie much below the *MR*R values at the $\pm$ 1 level. Apart from the curvature effect, one of the reasons for this could be the effect of changes in the pulse– off time which accompany the changes in pulse–on time. Since the duty factor is held constant during the experiments, the pulse–off time changes with changes in $T_{on}$. To investigate this further, experiments were conducted to observe the changes in *MR*R with $T_{off}$ at a constant value of $T_{on}$. The results are shown in Fig. 7. It can be seen that for a very low value of $T_{off}$, the *MR*R is low. As the $T_{off}$ is increased, *MR*R increases drastically and then falls down slowly with further increase in $T_{off}$. It was visually observed that vigorous sparking takes place for the $T_{off}$ conditions when *MR*R was very high, suggesting a different discharge mechanism. One of the probable causes could be the presence of an explosive mode as discussed by Kunieda *et al*. [4], which is active within a specific $T_{off}$ band. Thermally activated spontaneous oxidation reaction takes place in the explosive mode even during the pulse–off time leading to a relatively high *MRR*.

## 6 Conclusions

In the present work, experimentally obtained polynomial models for *MR*R and *R*a have been used for optimizing the dry EDM process. Process optimization has been performed using Non–dominated Sorting Genetic Algorithm (NSGA II) to obtain the Parto–optimal front. The resultant Pareto– optimal front can be used to select the appropriate operating conditions depending on the specific *MR*R or *R*a requirements. Based on the optimization results and the subsequent focused experiments, the following conclusions can be drawn:

> 1. Multi–objective optimization revealed that high air pressure and high spindle speed combination is favorable for obtaining both a high MRR and a low Ra. Such a combination of these input parameters leads to a higher flushing efficiency. Finish machining region (low Ra) was obtained for low current, high pulse–on time and low duty factor values. Rough machining region (high MRR) was obtained for high current, low pulse–on time and high duty factor values. Best Ra value of 1.60 µm and best MRR value of 6.83 mm3/min was obtained experimentally in the finish and rough regions, respectively.

2. Focused experiments conducted in the finish and rough machining regions revealed the existence of an additional process constraint based on the flushing efficiency. Flushing efficiency cannot increase indefinitely with an increase in air pressure and spindle speed since it saturates to a maximum value beyond some combination of air pressure and spindle speed. Thus, the optimization model presented here overestimates the MRR and underestimates the Ra value.

3. The air pressure and spindle speed values for flushing efficiency saturation depend on the amount of debris produced during machining and hence depend on the MRR. Under high MRR conditions, higher pressure and spindle speed values are required for obtaining the saturation point of flushing efficiency. Thus, the optimization results are closer to the experimental values in the rough machining region than in the finishing region. In the rough machining region, the optimum MRR is within 20% and Ra within 5% of the experimental values.

An extension of the current work could be the identification of practical constraints (such as flushing efficiency) in terms of the input variables. Incorporating such constraints into the empirical model would help in obtaining more realistic optimization results.

# Tables

Table 1: Parameter values corresponding to the coded levels for the CCD runs

| Coded Levels → <br> Parameter ↓ | $-\alpha$ | $-1$ | $0$ | $1$ | $+\alpha$ |
|---|---|---|---|---|---|
| Gap voltage, $V_g$ (V) | 55 | 63 | 77 | 91 | 99 |
| Discharge current, $I_d$ (A) | 9 | 16 | 29 | 42 | 49 |
| Pulse–on time, $T_{on}$ ($\mu s$) | 50 | 200 | 500 | 750 | 1000 |
| Duty factor, $D$ (%) | 8 | 24 | 48 | 72 | 88 |
| Air inlet pressure, $P$ (kPa) | 58.8 | 88.2 | 147 | 205.8 | 245 |
| Spindle speed, $N$ (rpm) | 300 | 650 | 1275 | 1900 | 2250 |

Table2: Regression statistics for the fitted models of MRR and Ra

| Model → <br> Statistics ↓ | $MRR$ | $Ra$ |
|---|---|---|
| $R^2$ | 0.907 | 0.64 |
| Adjusted $R^2$ | 0.892 | 0.592 |
| Predicted $R^2$ | 0.869 | 0.523 |

Table 3: A representative set of Pareto–optimal solutions and the corresponding input parameter values obtained after NSGA II run

| $MRR$ mm$^3$/min | $Ra$ $\mu$m | $V$ V | $I$ A | $T_{on}$ $\mu$s | $D$ % | $P$ kPa | $N$ rpm |
|---|---|---|---|---|---|---|---|
| 0.89 | 1.86 | 79.9 | 9.0 | 1000 | 8.0 | 245 | 2250 |
| 1.00 | 1.87 | 82.2 | 9.6 | 1000 | 8.0 | 245 | 2250 |
| 1.14 | 1.89 | 82.3 | 11.1 | 1000 | 8.0 | 245 | 2250 |
| 1.33 | 1.92 | 81.6 | 13.4 | 1000 | 8.0 | 245 | 2250 |
| 1.57 | 1.96 | 80.7 | 16.2 | 1000 | 8.0 | 245 | 2250 |
| 1.96 | 2.02 | 78.9 | 20.8 | 1000 | 8.0 | 245 | 2250 |
| 2.58 | 2.11 | 76.9 | 27.8 | 1000 | 8.0 | 245 | 2250 |
| 3.15 | 2.19 | 74.3 | 34.2 | 1000 | 8.0 | 245 | 2250 |
| 2.87 | 2.15 | 78.0 | 30.7 | 1000 | 8.0 | 245 | 2250 |
| 3.83 | 2.27 | 72.4 | 41.3 | 1000 | 8.0 | 245 | 2250 |
| 4.61 | 2.36 | 71.0 | 49.0 | 1000 | 9.6 | 245 | 2250 |
| 5.34 | 2.47 | 72.1 | 49.0 | 1000 | 42.9 | 245 | 2250 |
| 6.58 | 2.58 | 71.3 | 44.3 | 50 | 81.1 | 245 | 2250 |
| 8.02 | 2.72 | 61.4 | 49.0 | 50 | 88.0 | 245 | 2250 |
| 8.18 | 2.84 | 55.3 | 49.0 | 50 | 88.0 | 245 | 2250 |

Table 4: Pareto–optimal solutions belonging to the finish machining region

| $MRR$ mm$^3$/min | $Ra$ $\mu$m | $V_g$ V | $I_d$ A | $T_{on}$ $\mu$s | $D$ % | $P$ kPa | $N$ rpm |
|---|---|---|---|---|---|---|---|
| **0.886** | **1.857** | **80.0** | **9** | **1000** | **8.0** | **245** | **2250** |
| 0.886 | 1.857 | 80.0 | 9 | 1000 | 8.0 | 245 | 2250 |
| 0.895 | 1.857 | 80.3 | 9 | 1000 | 8.0 | 245 | 2250 |
| 0.897 | 1.858 | 80.4 | 9 | 1000 | 8.1 | 245 | 2250 |
| 0.905 | 1.858 | 80.6 | 9 | 1000 | 8.0 | 245 | 2250 |
| 0.906 | 1.858 | 80.7 | 9 | 1000 | 8.0 | 245 | 2250 |
| 0.910 | 1.858 | 80.8 | 9 | 1000 | 8.0 | 245 | 2250 |
| 0.912 | 1.858 | 80.9 | 9 | 1000 | 8.0 | 245 | 2250 |
| 0.914 | 1.858 | 80.9 | 9 | 1000 | 8.0 | 245 | 2250 |
| 0.920 | 1.859 | 81.2 | 9 | 1000 | 8.0 | 245 | 2250 |

Table 5: Experimental observations for CCD runs in the finish machining region; $T_{on}$ = 1000 µs, D = 8%, P = 245 kPa, N = 2250 rpm

| Run Order | $V_g$ V | $I_d$ A | MRR mm³/min | Ra µm |
|---|---|---|---|---|
| 1 | 82 | 7 | 0.26 | 2.25 |
| 2 | 75 | 7 | 0.27 | 2.15 |
| 3 | 75 | 4 | 0.11 | 1.88 |
| 4 | 75 | 10 | 0.45 | 2.34 |
| 5 | 68 | 7 | 0.26 | 1.6 |
| **6** | **80** | **9** | **0.37** | **2.21** |
| 7 | 75 | 7 | 0.24 | 2.42 |
| 8 | 75 | 7 | 0.24 | 1.79 |
| 9 | 80 | 5 | 0.14 | 1.79 |
| 10 | 75 | 7 | 0.22 | 2.01 |
| 11 | 70 | 5 | 0.1 | 1.75 |
| 12 | 70 | 9 | 0.29 | 1.94 |
| 13 | 75 | 7 | 0.21 | 1.83 |

Table 6: Pareto–optimal solutions belonging to the rough machining region

| MRR mm³/min | Ra µm | $V_g$ V | $I_d$ A | $T_{on}$ µs | D % | P kPa | N rpm |
|---|---|---|---|---|---|---|---|
| **8.176** | **2.844** | **55.3** | **49** | **50** | **88.0** | **245** | **2250** |
| 8.175 | 2.843 | 55.3 | 49 | 50 | 88.0 | 245 | 2250 |
| 8.167 | 2.838 | 55.5 | 49 | 50 | 88.0 | 245 | 2250 |
| 8.167 | 2.838 | 55.5 | 49 | 50 | 88.0 | 245 | 2250 |
| 8.161 | 2.832 | 55.7 | 49 | 50 | 87.9 | 245 | 2250 |
| 8.161 | 2.832 | 55.7 | 49 | 50 | 87.9 | 245 | 2250 |
| 8.155 | 2.826 | 56.0 | 49 | 50 | 87.9 | 245 | 2250 |
| 8.146 | 2.818 | 56.3 | 49 | 50 | 87.9 | 245 | 2250 |
| 8.145 | 2.816 | 56.4 | 49 | 50 | 88.0 | 245 | 2250 |
| 8.130 | 2.807 | 56.8 | 49 | 50 | 87.9 | 245 | 2250 |

Table 7: Experimental observations for factorial and center runs in the rough machining region; $I_d$ = 49 A, D = 88%, P = 245 kPa, N = 2250 rpm

| Run Order | $V_g$ V | $T_{on}$ μs | MRR mm³/min | Ra μm |
|---|---|---|---|---|
| 1 | 50 | 30 | 2.08 | 3.53 |
| 2 | 50 | 30 | 2.82 | 3.14 |
| 3 | 45 | 50 | 5.06 | 3.71 |
| 4 | 55 | 10 | 3.26 | 3.51 |
| 5 | 45 | 10 | 2.56 | 3.05 |
| **6** | **55** | **50** | **6.83** | **2.97** |
| 7 | 50 | 30 | 1.49 | 3.27 |

# Figure Captions

Figure 1: Schematic of dry EDM process

Figure 2: Flowchart for NSGA II algorithm

Figure 3: Pareto-optimal solutions obtained after the NSGA II run

Figure 4: Effect of (a) gap voltage and (b) discharge current, on $R$a in finish machining region

Figure 5: Interaction plot for $R$a in finish machining region

Figure 6: Effect of (a) gap voltage and (b) pulse–on time, on $MR$R in rough machining region

Figure 7: Effect of pulse–off time on $MR$R in rough machining region

Figure 1

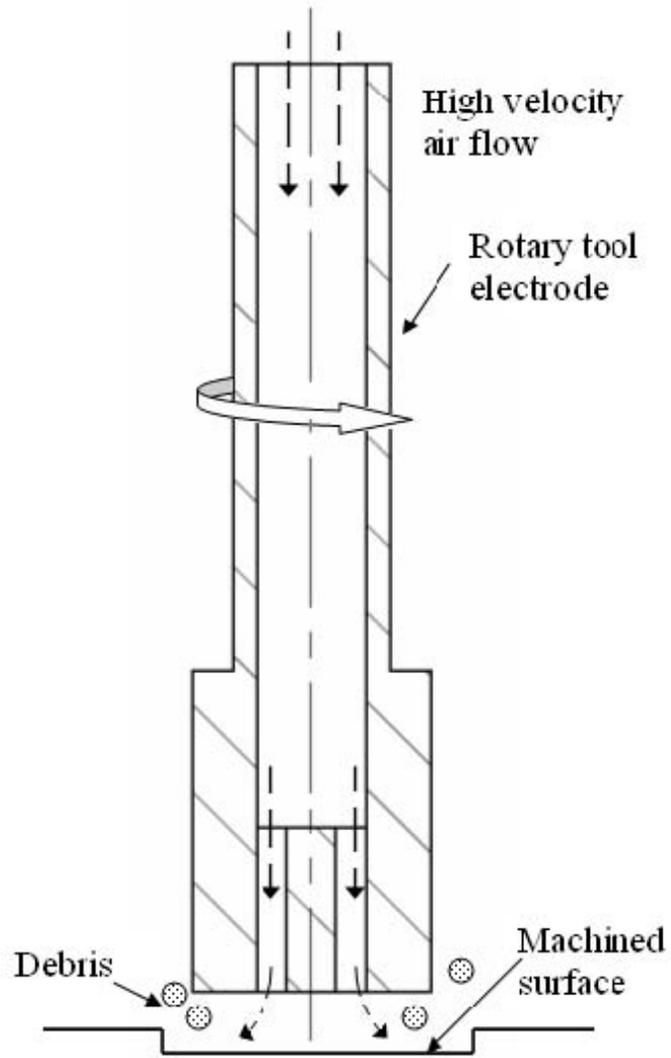

Figure 2

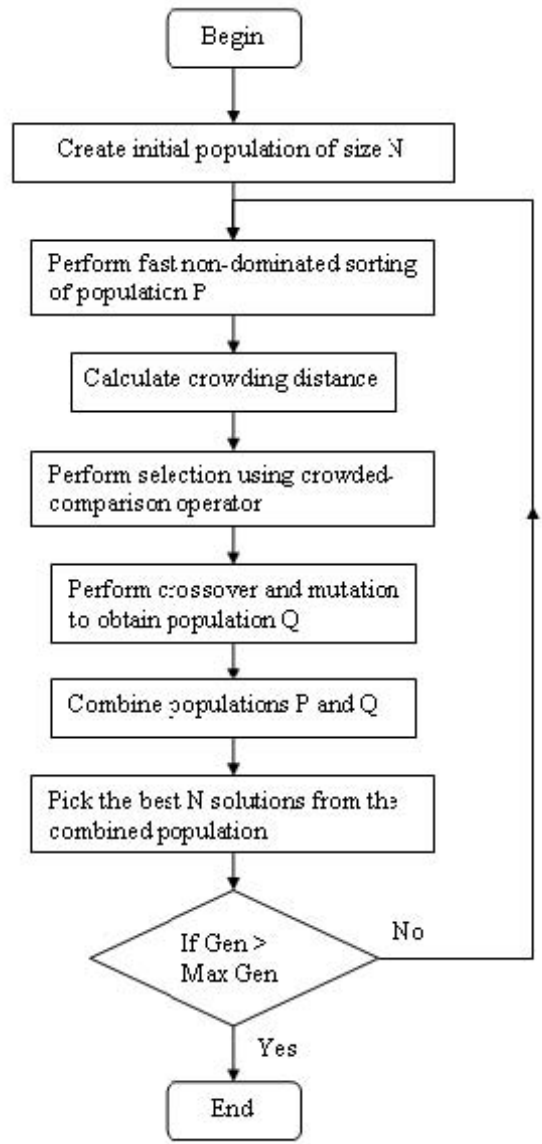

Figure 3

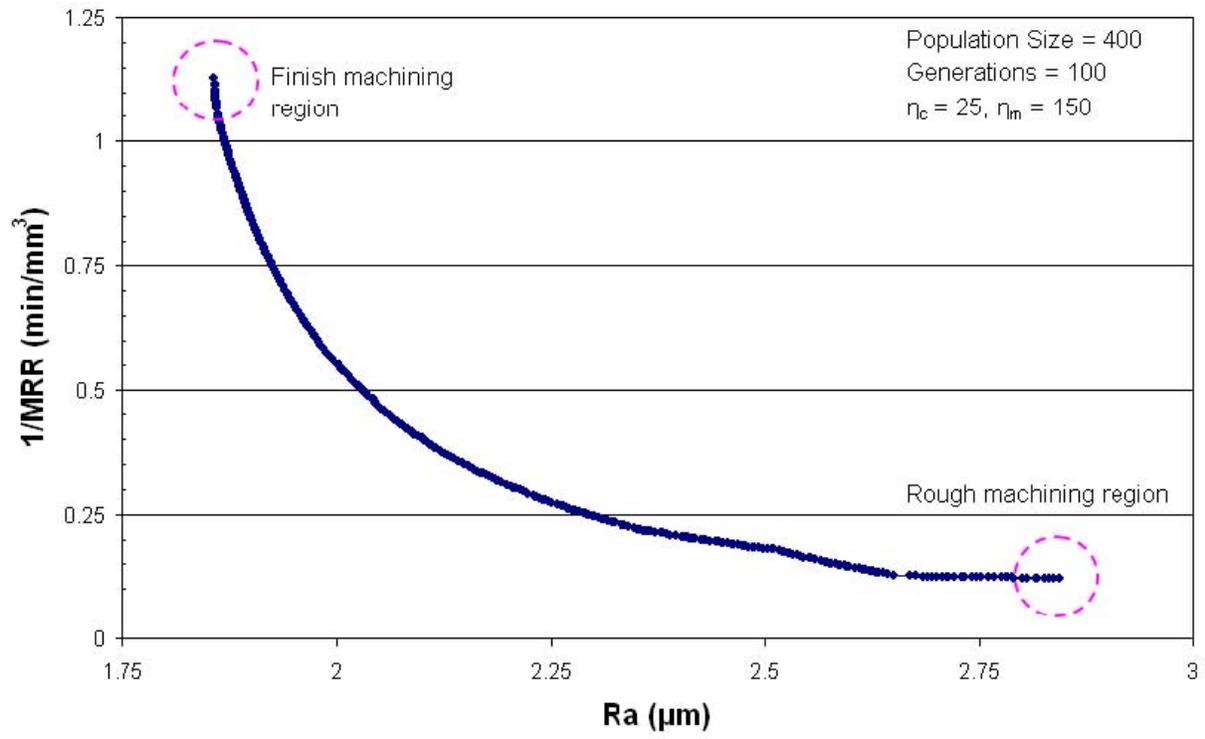

Figure 4 (a)

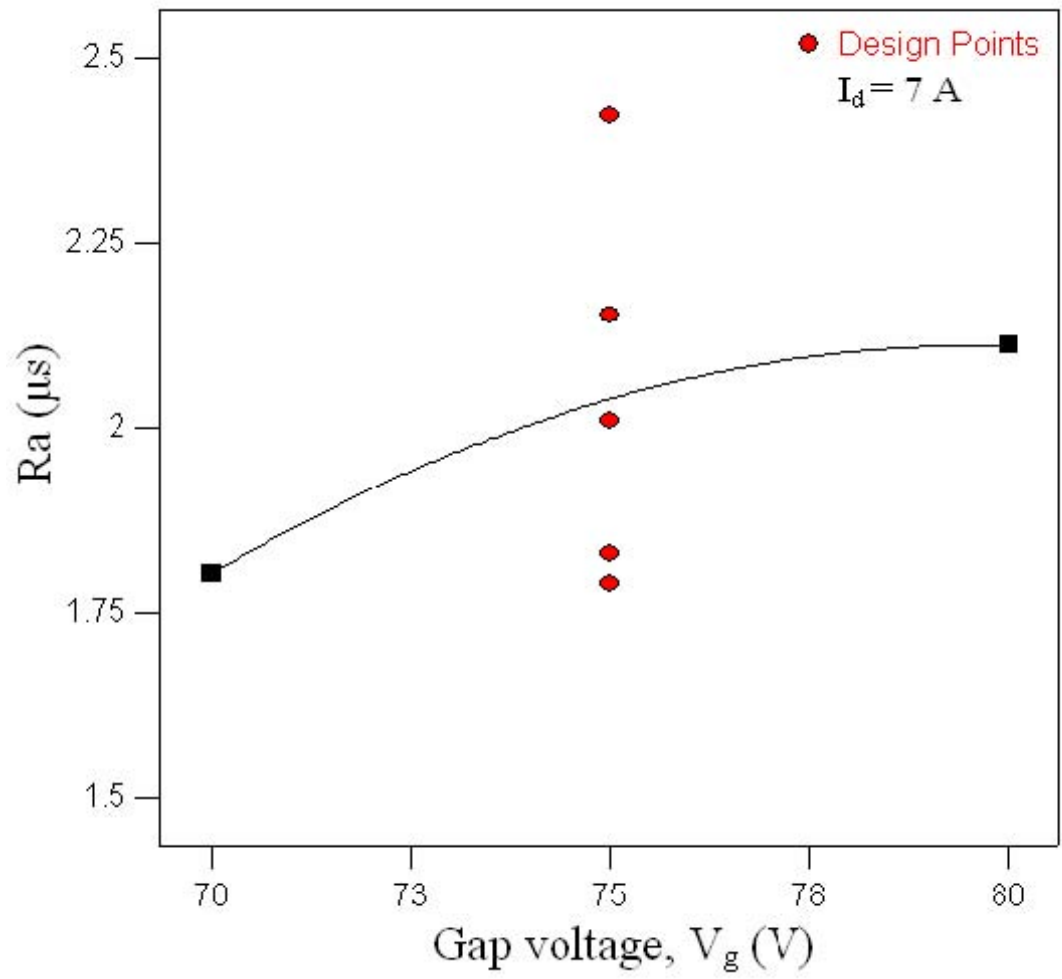

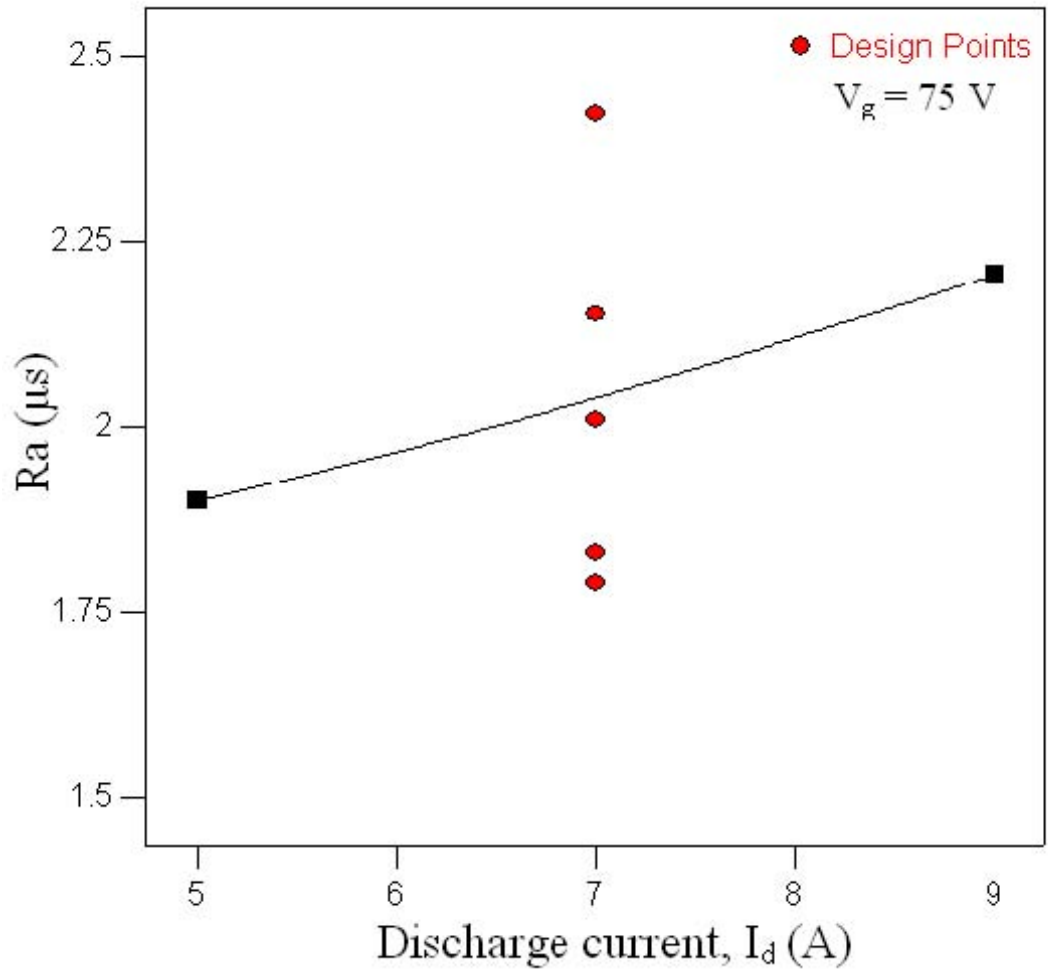

Figure 4 (b)

Figure 5

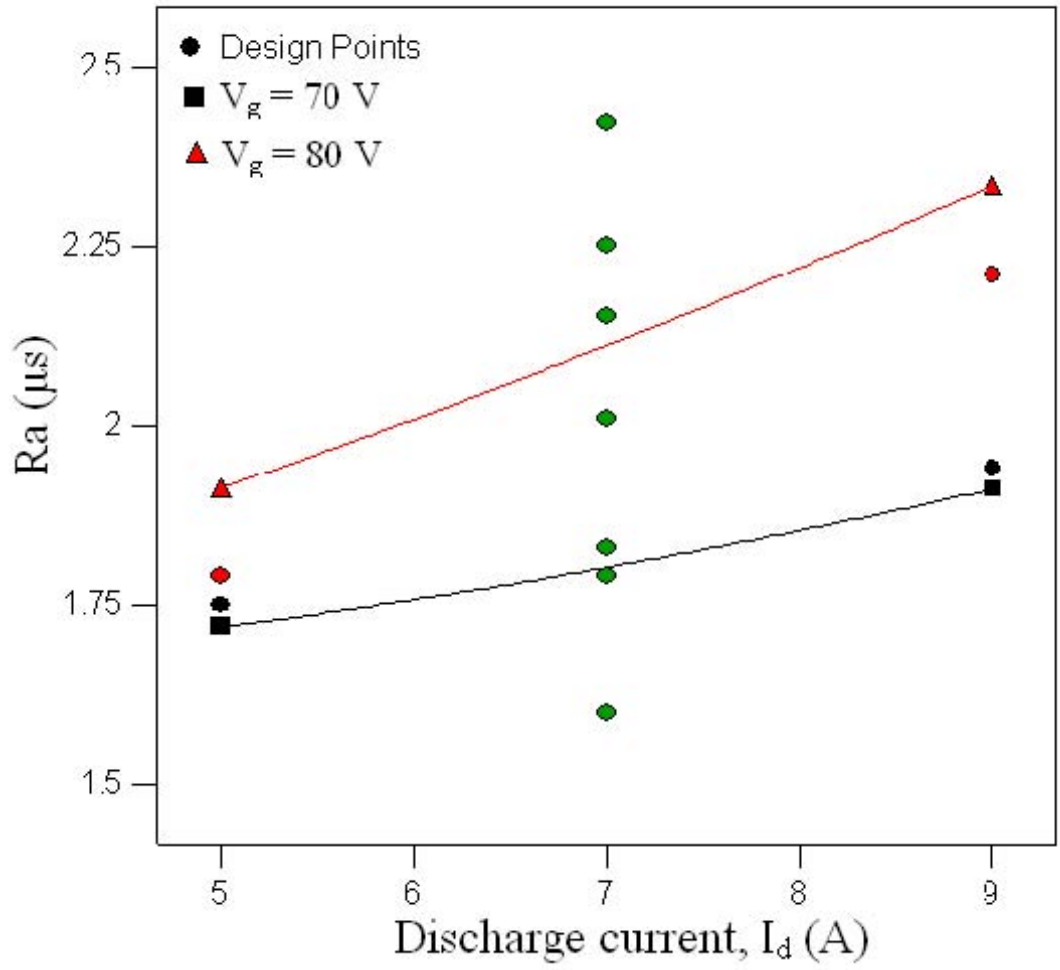

Figure 6 (a)

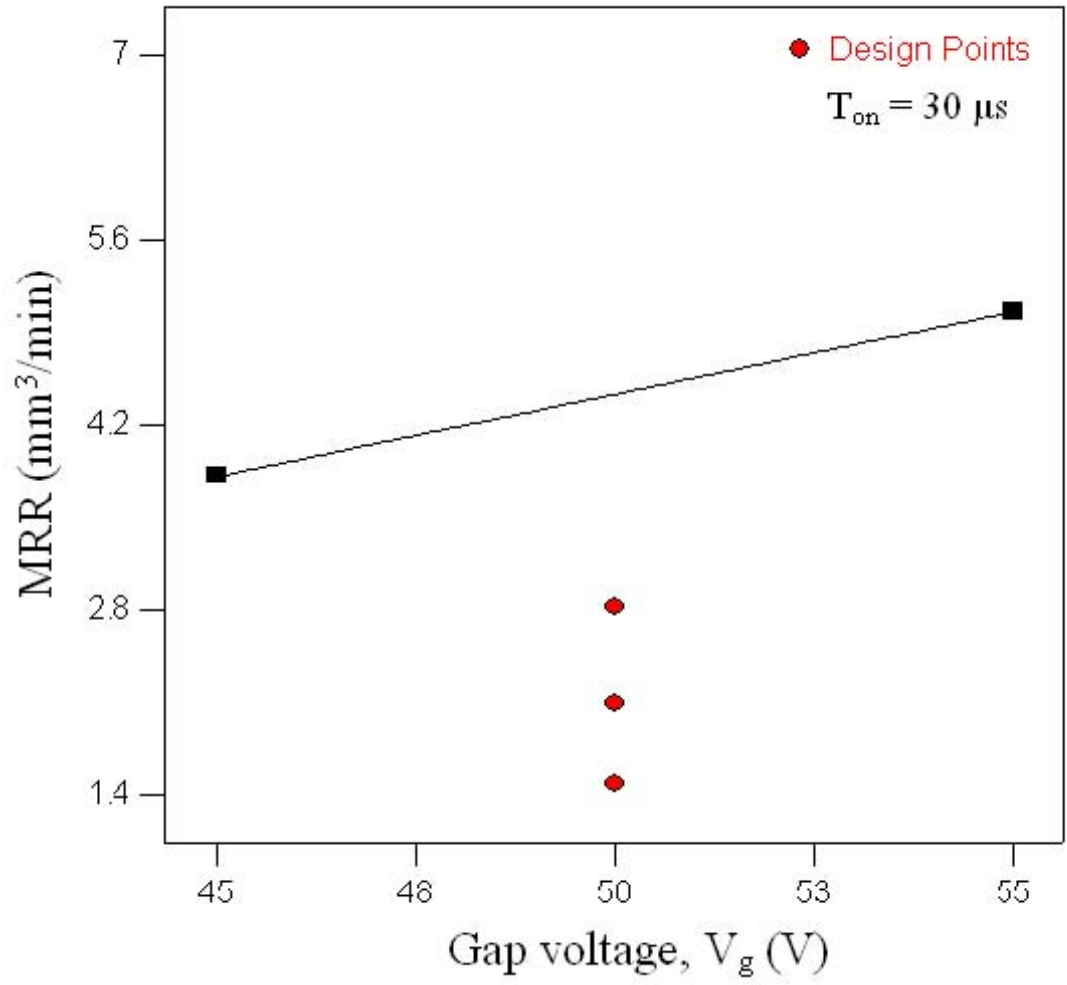

Figure 6 (b)

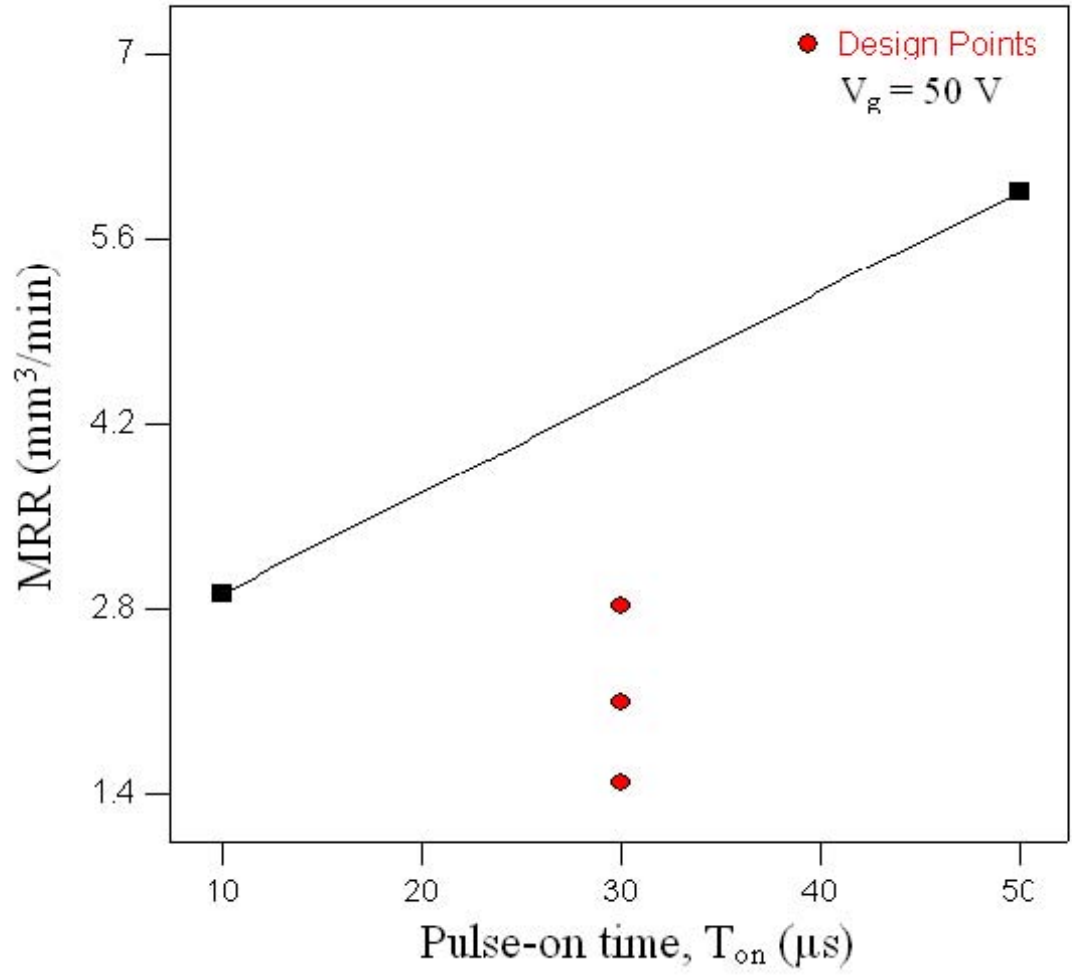

Figure 7

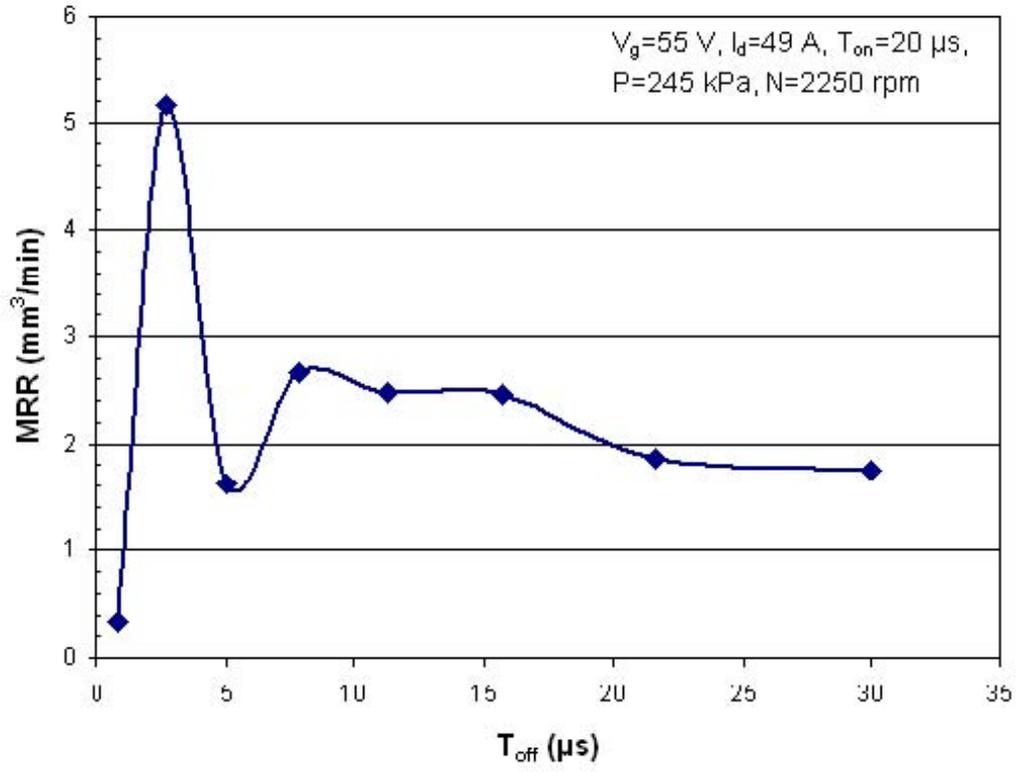